\documentclass[twocolumn]{article}
\usepackage[english]{babel}

\usepackage[affil-it]{authblk}
\usepackage{amsfonts}
\usepackage{amssymb}
\usepackage{amsthm}
\usepackage{amsmath}

\newtheorem{theorem}{Theorem}[section]

\newtheorem{lemma}[theorem]{Lemma}

\usepackage{braket}
\usepackage{verbatim}
\usepackage{listings}
\usepackage{rotating}
\usepackage{hhline}
\usepackage[table,x11names]{xcolor} % colour lines in table, not beamer compatible
\definecolor{vlightgray}{cmyk}{0,0,0,0.125}

\usepackage{tikz}
\usetikzlibrary{arrows.meta}
\usetikzlibrary{quantikz}

\usepackage[letterpaper,top=2cm,bottom=2cm,left=3cm,right=3cm,marginparwidth=1.75cm]{geometry}

\usepackage{amsmath}
\usepackage{graphicx}
\usepackage[colorlinks=true, allcolors=blue]{hyperref}

\title{On Zero-Knowledge Proofs over the Quantum Internet}
\author[1]{Mark Carney}
\affil[1]{Quantum Village Inc.}
\date{February 2023}

\begin{document}

%\twocolumn[
%\begin{@twocolumnfalse}
\maketitle

\begin{abstract}
This paper presents a new method for quantum identity authentication (QIA) protocols. The logic of classical zero-knowledge proofs (ZKPs) due to Schnorr \cite{Schnorr89} is applied in quantum circuits and algorithms. This novel approach gives an exact way with which a prover $P$ can prove they know some secret by encapsulating it in a quantum state before sending to a verifier $V$ by means of a quantum channel - allowing for a ZKP wherein an eavesdropper or manipulation can be detected with a fail-safe design. This is achieved by moving away from the hardness of the Discrete Logarithm Problem towards the hardness of estimating quantum states. This paper presents a method with which this can be achieved and some bounds for the security of the protocol provided. With the anticipated advent of a `quantum internet', such protocols and ideas may soon have utility and execution in the real world. 
\end{abstract}
%\end{@twocolumnfalse}]

\section{Introduction}

With the advent of Quantum Computing comes with it the idea of the Quantum Internet - the ability to transfer a quantum state $\ket{\Psi}$ from one quantum computer/device to another. There are many challenges with this kind of networking \cite{Cacciapuoti2020}, as well as many benefits. As Cacciapuoti \cite{Cacciapuoti2020} points out, with a quantum internet we get Quantum Key Distribution `for free', a major benefit to quantum communications infrastructure. There are many existing Quantum Identity Authentication (QIA) protocols \cite{Dutta2021} and this paper adds a new approach to the collection.

Existing approaches make use of various features of QKD, quantum teleportation techniques, Physically Unclonable Functions (PUFs), distributed Bell states, quantum private queries, quantum secure direct communications, etc. Many of these details may be found in \cite{Dutta2021}.

Schnorr introduced in \cite{Schnorr89} the idea of efficient identification signatures, initially designed for use with smart cards. This method of `proving' your identity without disclosing a secret became known as `zero-knowledge proofs' and have recently found much use in many cryptographic protocols \cite{Morais2019}.

The benefits of ZKPs over other past approaches are that there needs be no prior exchange or other pre-sharing, nor any explicit statement of what the hidden information is. The proof system itself carries the correctness and soundness that guarantees the validity of a proof presented by the prover to the verifier, and that the claim by the prover to know such a secret is `true'.

ZKPs have been used to create quantum proof systems that have also been shown to be possible in a quantum setting \cite{Watrous05}. These make use of graph isomorphism problems, which this approach does not. The  method herein takes advantage of a quantum communications network to reduce the number of quantum and classical transmissions down to four and three respectively. 

The work presented here aims to demonstrate how a quantum ZKP protocol might look by coding Schnorr's original method into quantum states. Some benefits and restrictions of this approach are included. 

\section{Schnorr ZKP Protocol}

In its simplest form, a zero-knowledge proof is a method for a prover $P$ to provide a way of showing that they know some secret $x$ to a verifier $V$, but without exposing the secret at any point, hence `zero-knowledge'. 

The following algorithm is the usual presentation of Schnorr's work. $P$ wants to prove that they know $x$ such that $Y = g^x \mod p$, for prime $p$ and generator $g$, with $g$, $p$, and $Y$ public. The following method is presented:
\begin{enumerate}
    \item \underline{$P \rightarrow V$:} $P$ chooses some $r$ and sends ${t = g^r \mod p}$ to $V$.
    \item \underline{$V \rightarrow P$:} $V$ sends a random $c$ to $P$.
    \item \underline{$P \rightarrow V$:} $P$ sends $s = r + cx$ to $V$
    \item \underline{$V$ checks} that $g^s \equiv t \times Y^c \mod p$.
\end{enumerate}

This works as
\begin{equation}
\begin{aligned} \label{eq:schnorr}
%\begin{split}
t \times Y^c & \equiv g^r \times (g^x)^c & \mod p \\
& \equiv g^{r+cx} & \mod p \\
& \equiv g^s & \mod p
%\end{split}
\end{aligned}
\end{equation}

This very neat scheme was a very important development in authentication schemes, and will form the basis for the quantum protocol presented next.

\section{Quantum Preliminaries}

This protocol utilises a single qubit, and only two quantum gates. Qubits are assumed to be initialised in $\ket{0} = \begin{psmallmatrix} 1 \\ 0 \end{psmallmatrix}$ with our target state $\ket{1} = \begin{psmallmatrix} 0 \\ 1 \end{psmallmatrix}$. With $\alpha, \beta \in \mathbb{C}$, $\ket{\psi} = \begin{psmallmatrix} \alpha \\ \beta \end{psmallmatrix}$, such that $|\alpha|^2 + |\beta|^2 = 1$. Quantum circuits are formed from products and tensor products of $2 \times 2$ unitary matrices, referred to as quantum gates (analogous to binary gates), preserving the unitary property \cite{Nielsen2010}. 

Define the $R_x$ gate as \cite{Nielsen2010}:
\begin{align}\label{eq:rgate}
\begin{split}
R_x(\theta) & = e^{i \theta X / 2} \\
& = \cos{(\theta/2)}I + i \sin{(\theta/2)}X \\
& = \begin{pmatrix}
    \cos(\theta/2) & -i \sin(\theta/2) \\
    - i \sin(\theta/2) & \cos(\theta/2)
\end{pmatrix}
\end{split}
\end{align}
where $I = \begin{psmallmatrix} 1 & 0 \\ 0 & 1 \end{psmallmatrix}$ and $X = \begin{psmallmatrix} 0 & 1 \\ 1 & 0 \end{psmallmatrix}$. With the representation of the Bloch sphere, this gate is usually interpreted as a rotation along the $x$ axis. %We shall also require $Y = \begin{psmallmatrix} 0 & -i \\ i & 0 \end{psmallmatrix}$ and $Z = \begin{psmallmatrix} 1 & 0 \\ 0 & -1 \end{psmallmatrix}$, interpreted as $y$ and $z$ axis rotations respectively.

The following gates $G_p(a)$ and $H_p(a)$ shall be utilised, defined as follows:
\begin{align}\label{eq:ggate}
G_p(a) = R_x \Big((a \mod p) \times \frac{\pi}{p} \Big) \\
H_p(a) = R_x \Big((a \mod 2p) \times \frac{\pi}{p} \Big)
\end{align}

Intuitively, we split the $\pi$ rotation about the $x$ axis on the Bloch sphere into $p$ many steps, and then apply a rotation on our qubit, moving that number of steps around. The important thing to note here is that $G_p(a) G_p(b) = H_{p}(a+b)$, which can be made $G_p(a + b)$ by applying $X$ if $(a + b \mod 2p) > p$. This will be useful later.%The other $Y$ and $Z$ gates will be used to correct the rotations later should it be required.

%Let $$\ket{G_p(g^a)}^n = \underbrace{\ket{G_p(g^a)} \ket{G_p(g^a)} \ldots \ket{G_p(g^a)}}_{\text{$n$-many times}} $$

Let $k_p(n)$ be defined as \begin{align}  k_p(n) = \begin{cases} 0 & \text{ if } ( n \mod 2p) < p \\ 1 & \text{ otherwise } \end{cases}\end{align} and let $C_{m} = X$ gate if $m = 1$, else $C_m = I$. 

\section{Quantum Internet ZKPs}\label{sec:QZKP}

This section brings these two domains together to propose an authentication scheme that makes use of a quantum internet with additional classical channel.

\subsection{Q-ZKP Protocol}

The Quantum Internet, loosely defined, is a quantum communications protocol that permits the transfer of some quantum state $\ket{\Psi}$ from one quantum computer/device to another. Utilising this property, the following zero-knowledge proof can be constructed. 

As before, $P$ wishes to prove they know $x$ to $V$, in this case such that they can create a state $G_p(x)\ket{0}$. Both the gate $G_p$ and value of $p$ are known publicly. 

\begin{enumerate}
    \item \underline{$V$ selects} random values  $c$ and $n$.
    \item  \underline{$V \rightarrow P$:} Let $V$ have $\ket{x}=G_p(x)\ket{0}$, but no knowledge of $x$. $V$ sends to $P$ \begin{align} \ket{x+(c-1)n} = G_p((c-1)n)\ket{x}\end{align}
    \item \underline{$P \rightarrow V$} $P$ selects some random $r$ and sends the state: \begin{align}\label{eqn:A} \ket{A} = G_p(r)\ket{x + (c-1)n} \end{align}
    \item \underline{$V \rightarrow P$:} $V$ sends $c$ over a classical channel and sends the state $$\ket{S_1} = G_p(n)\ket{A}$$
    \item \underline{$P$ computes}  $s = r + cx $. Let $  b = k_p(t) $ where \begin{align*}t = \Big(&(x \text{ mod } p)  + (r \text{ mod } p) \\ &  + (x(c-1) \text{ mod } p) \Big)\end{align*}
    \item \underline{$P \rightarrow V$:} $P$ sends $s$ and $b$ and then sends the state: $$\ket{S_2} = G_p(x(c-1))\ket{S_1}$$
    \item \underline{$V$ constructs} \begin{align}\label{eq:B} \begin{split} \ket{B} &= G_p(-cn)\ket{S_2 } \end{split}\end{align} and calculates \begin{align}\begin{split} a &= k_p \Big( ((c-1)n \text{ mod } p) + \\ & (n \text{ mod } p)  + (-cn \text{ mod } p) \Big) \end{split}\end{align}
    \item \underline{$V$ checks} that $$ G_p(p-s) C_{a \oplus b} \ket{B} = \ket{1} $$ by seeking a $1$ under the normal $z$ axis measurement.
\end{enumerate}

\subsubsection{Note on Notation}

It should be made clear that the various states are applied successively to received states. Whilst combining rotations from distinct states is hard, applying rotations to received states is straightforward theoretically, especially for commutative gates that are in use here. Following, for example, a quantum teleportation operation receiving state $\ket{\Psi}$, we apply gate $G_1$ then $G_2$ to obtain $G_2 G_1 \ket{\Psi}$.

\subsection{Correctness and Completeness}

\begin{lemma}\label{lemma:commute}
\begin{align}\label{eq:rxab} R_x(b)R_x(a)\ket{0} = R_x(a+b)\ket{0} \end{align}
\end{lemma}

\begin{proof}
\begin{align*}
R_x (a)\ket{0} &= \begin{pmatrix} cos(\frac{a}{2}) \\ - i sin(\frac{a}{2}) \end{pmatrix} = \ket{a} \\
R_x(b)\ket{a} &= \begin{pmatrix} cos(\frac{a}{2})cos(\frac{b}{2}) - sin(\frac{a}{2})sin(\frac{b}{2}) \\ -i(cos(\frac{a}{2})sin(\frac{b}{2}) + sin(\frac{a}{2})cos(\frac{b}{2}) ) \end{pmatrix} \\
 &= \begin{pmatrix} cos(\frac{a+b}{2}) \\ -i sin(\frac{a+b}{2}) \end{pmatrix} \\
  &= R_x(a+b)
\end{align*}
\end{proof}

From this follows also the commutativity of single axis rotations \begin{align} R_x(a)R_x(b) = R_x(b)R_x(a) \end{align} It then further follows that in equation (\ref{eq:B}) \begin{align} G_{p}(x(c-1))\ket{A} = G_p(r) H_{p}(xc) \ket{0} \end{align}

Next we need to take $H_{p}(r+xc)$ which is formed from full rotations about the $x$ axis, and restrict it down to half-axis rotations. This is where $C_b$ comes in to play. 

Note that if some $$(a \mod 2p) > p$$ then $$(a + p \mod 2p) < p$$ Given our $X$ gate effectively fulfils this function, it is conditional on $P$'s assessment in witness $b$ whether it is applied or not. As such if $$(r+xc \mod 2p) > p$$ then \begin{align}\label{eqn:x-norm}  X H_{p}(xc) G_p(r) =  G_p(xc)G_p(r)\end{align}

This gives us, given a correct choice of $C_b$ \begin{align}\label{eq:BC} B C_b = G_p(xc+r) = G_p(s)\end{align} We use this for the $cn$ construction also, noting that if both overflow then we need do nothing, and so use the XOR of our two evaluations as two overflows do not need correcting.

We then need the following theorem to complete our proof's validity:

\begin{theorem}
    Let $C_{a \oplus b}$ be chosen appropriately as above. When $V$ implements the protocol as outlined above the output will always be a $\ket{1}$ if and only if $V$ agrees that $P$ has a valid proof that they know $x$.
\end{theorem}

\begin{proof}
$(\leftarrow)$ Start by re-asserting the interpretation of equation (\ref{eq:schnorr}) in this scheme, namely that for a valid proof it follows that  $$ s \equiv r + xc \mod p $$

By Lemma \ref{lemma:commute} and equation (\ref{eqn:x-norm}), $$ C_a G_p(-cn)G_p(n)G_p((c-1)n) = G_p(0) $$ It then follows that, equations (\ref{eqn:A}) and (\ref{eq:BC}):
\begin{align}\label{eqn:main}
\begin{split}
G_p & (p-s)C_bG_p(x(c-1))G_p(x+r)\ket{0} \\
&= G_p(p-(r+xc))G_p(r+xc) \ket{0} \\
&= G_p(p) = R_x(\pi)\ket{0} = \ket{1}
\end{split}
\end{align}

Given the protocol only divides a half, not a full, qubit rotation by $p$ this completion should always send the qubit to be in state $\ket{1}$.

Therefore have a $1$ measurement, modulo some error $\epsilon$. 

$(\rightarrow)$ If the measurement output is (almost) always $1$ modulo some noise, then the states received from $P$ by $V$ matches the $\ket{1}$ state expected by $V$.

By equation (\ref{eqn:main}) a $\ket{1}$ state, and subsequent $1$ measurement means that everything required to line up in this scheme has done so, and $P$'s proof is correct.
\end{proof}

\subsection{Security}\label{sec:security}

If we paraphrase Shannon \cite{Shannon1949}, a perfectly secure zero-knowledge proof is one in which the information disclosed about $x$ in a proof $P$ is null, essentially $$ I(x; P) = 0 $$

Whilst it may be correct for $V$ to arrive at a $\ket{1}$ state, there are several considerations that make sure that only a party $P$ who can generate a valid state $\ket{x} = G_p(x)\ket{0}$ can successfully complete a proof and satisfy verifier $V$, and that some attacker/eavesdropper cannot either discern the value of $x$ nor impersonate $P$ maliciously by inserting themselves midway through a ZKP sequence.

With the advent of Shor's algorithm (see \cite[Appendix 4]{Nielsen2010}) it is clear that for the classical scheme due to Schnorr, if $Y = g^x \mod p$ is public alongside $g$ and $p$, then $x$ may be recovered by means of this algorithm. As such, a way of sharing quantum states that encode $x$ and the subsequent proof is needed, which this protocol attempts to provide. 

To do this we substitute exponents over some $g$ for rotations about the $x$ axis on a qubit, relying on the hardness of decoding quantum states rather than the discrete logarithm problem.

There are two sides to this proof scheme's security; a classical side and a quantum one. Let $E$ denote some attacker/eavesdropper.

\subsubsection{Classical Security}

The classical security concerns the classical channels, and we assume some eavesdropper on these. The variables $c$ and $s$ can be publicly disclosed, as knowing $c$ does not help you in discerning the secret $x$ given the additive $r$ that is used.

\begin{theorem}\label{thm:classical}
    The classical security of the variables $x$, $r$, and $n$ is that an attacker $E$ has at most a $1/p$ chance to provide a malicious proof.
\end{theorem}

\begin{proof}
There are three options classically for an attacker to try and pursue when attacking this protocol:
\begin{enumerate}
    \item \textbf{Guess $x$} - this would be the most direct method, and would correctly compromise any proof from $P$. Given $p$ is prime, there are $p$ many options for both values but they can be checked against $s$ given $c$ is public. Therefore the probability of this occurring would be $1/p$. 
    \item \textbf{Guess $n$} - Focusing again on a PitM attack taking place after the initial state $\ket{A}$ was sent from $P$; $E$ does not know $n$ as it is multplied by $c$, and so cannot simply subtract the value. $E$ could guess the value for $n$, then apply the following attack to defeat the proof:
    \begin{enumerate}
        \item The attacker knows $c$ and correctly guesses $n$, chooses some $t$ and then sends to $V$: $$ \ket{S_2} = G_p(t)G_p(cn)\ket{0} $$
        \item Set $s = t$ and $b=0$, which completes a valid proof.
    \end{enumerate}
    There is a $1/p$ chance that this works.
    \item \textbf{Guess $r$} - An unlikely attack, this would compromise the proof but only for one instance, and is only effective if the attacker graduates from eavesdropper to an active person-in-the-middle attack (PitM). As above, the likelihood this works is $1/p$.
\end{enumerate}

By this argument, $E$ has at best a $1/p$ chance to guess a value that could allow them to provide a valid malicious proof.
\end{proof}

\subsubsection{Quantum Security}

Next follows the analysis of the security of this system over noisy quantum channels - both error corrected and not. 

\textbf{I. Error Corrected Case} 

If we first assume an error corrected channel (\emph{e.g.} using a scheme found in Calderbank and Shor \cite{Shor1996}) then the security relies upon the fact that states are only transmitted once. As such, an attacker having to measure say $\ket{A}$ multiple times in order to produce any kind of valid amplitude estimation, \emph{e.g.} in \cite{QAAE}, becomes a very difficult attack vector. Given each value is only transmitted at most once in its original state such an attack is not viable, and so $E$ would likely not attempt to carry it out.

Therefore the security in this case falls back to the classical case above.

\textbf{II. The Noisy Case}

For a given quantum channel that has noise, the probability that a qubit is successfully transmitted is $1-\epsilon$, for some (hopefully) small error term $\epsilon$. Whilst an attacker $E$ listening in on the channel will raise the noise by means of incorrect guesses and interference, these may be detected by comparing the number of 0 measurements with an accepted bound given by the error. If we let $P_{valid}$ represent a valid proof from the protocol in section \ref{sec:QZKP}, the fidelity of the protocol with noise can be characterised as the expectation $$ E\bra{1} P_{valid} \ket{0} = 1 - \epsilon $$

Note that the usual convention of talking about bit errors in our rotations does not apply necessarily to our axis rotation inputs $x$, $r$, $n$, or $s$. This is because an error of $G_p(a \pm 2^{w+1})$ would be considerably more noticeable from $G_p(a \pm 2^{w})$ for most choices of $w$. Therefore we can assume that any channel noise will largely only affect the least significant bits of our single qubit rotation parameters.

\begin{theorem}\label{thm:quantum}
    Let $p$ be given, and let the quantum channel error term $\epsilon = 1/e$, then there is at best a $$ \frac{1}{p} + \frac{2p}{e^2} $$ chance that an attacker $E$ can successfully pass an incorrect proof as a valid one to $V$ in the scheme above.
\end{theorem}

\begin{proof}
To begin with let $$\frac{1}{e} \geq \frac{1}{p}$$ Taking the most likely attack scenario in theorem \ref{thm:classical}, we may reason as follows; Suppose an attacker makes a close guess $cn_{guess} = cn \pm 1$, the resulting error in the final sum in equation \ref{eqn:main} with noise $\epsilon$ will give measurement expectations of \begin{align}\begin{split} E \bra{1} G_p(p) \ket{0} &= 1-\epsilon \\ &\equiv E\bra{1} R_x((p \pm 1)\pi/p) \ket{0} \end{split}\end{align} With the error in the channel as above then this would not be distinguishable from the value of $cn$ transmitted with noise.

Therefore the likelihood that the attacker chooses $n_{guess}$ that is close enough to $n$ to be masked by noise and thereby have a successful attack to give a malicious valid proof $P_{valid}$ is the same as choosing $n$ with no error ($1/p$) or making one of two valid choices from $\{cn-1, cn+1\}$ with noise masking it: \begin{align*} Pr(P_{valid} | cn_{guess}) &= \frac{1}{p} + \frac{2}{e} = \frac{2p+e}{pe} \end{align*} For $e$ close to $p$, this would be around $3/p$, which is what should be expected.

However, in general we may find that $e > p$. Thereby this $2/e$ term decomposes as two instances of the proportion of $p$ to $e$ over $e$, or $\frac{p}{e^2}$; one for the likelihood of $E$ guessing $n+1$ and one for guessing $n-1$. This gives a combined upper bound of \begin{align}\begin{split} Pr(P_{valid} | n_{guess}) &= \Big( \frac{1}{p} + \frac{2p}{e^2} \Big) \end{split}\end{align}

%$$ \frac{3}{p} + \frac{1}{e} = \Big( \frac{3e+p}{ep} \Big) $$

%It follows that with this value of $\epsilon$ each exchange will increase the error by $1/p$, giving a total worst case error of $G_p(\pm 3)$ over the entire proof in equation (\ref{eqn:main}). Given each error is independent, there is a $1/2p$ chance that each error will occur with the correct choice between plus/minus 1 on the rotation. Given this occurs 3 times over 3 quantum exchanges, the overall probability is $(1/2p)^3 = 1/8p^3$.
\end{proof}

Note that because the `guess $n$' attack only affects one quantum transmission, we only need to consider the error once.

\subsubsection{Considerations Within the Protocol}

There are a number of security considerations within the protocol that we will state here. 

With the communications being hybrid classical and quantum, so is our `challenge'. Thereby we have to values, $c$ and $n$ that are both used in tandem to provide the challenge to $P$ that can only be resolved if $P$ knows $x$. To prevent $P$ disregarding the $\ket{x}$ that $V$ has, this challenge is commited to at the start of the protocol, and unwound fully at the end.

The choice of $r$ is never transmitted classically, and so is totally unknown to $V$. Likewise, $n$ is totally unknown to $P$, and even if $P$ is malicious they cannot unwind $G_p(cn)$ as $P$ is unaware how many times to apply $G_p(-c)$ as they do not know $n$. 

Therefore by delivering $G_p((c-1)n)$ at first, $P$ cannot simply prove they know \emph{any} $x$, just specifically the one that $V$ has in $\ket{x}$ at the start. Thereby, whilst $r$ creates a lock on this particular proof for $P$, $c$ and $n$ create a hybrid quantum-classical zero-knowledge challenge for $P$ to provide a resolution to.

It should be noted that if $V$ has some gate $U_x$ such that, without knowing $x$, $V$ may obtain $$U_x\ket{0} = G_p(x)\ket{0}$$ then some steps in the protocol become unnecessary, as $V$ can just construct $G_p(xc)$ themselves - they only need to receive state $\ket{A}$ and $s$. The author is, however, unaware of how this could be achieved without falling afoul to a protracted quantum amplitude estimation attack, for example.

\subsubsection{Overall Security}

The attack likelihood given in theorem (\ref{thm:quantum}) is the combined `worst-case' scenario for the protocol presented in this paper. 

By theorem (\ref{thm:quantum}) as $p$ increases and/or $\epsilon$ decreases then the number of repetitions required to validate a proof decreases according to the required confidence level. 

For a $5\sigma$ confidence, with an additive noise error of $\epsilon$ as defined in theorem \ref{thm:quantum} we would need $N$-many iterations such that $$ \Big( \frac{1}{p} + \frac{2p}{e^2} \Big)^N < 5.733 \times 10^{-7} $$ This would give us the highest confidence that $P$ was both honest and knew a value for $x$.

By analysing the effects of noise and how an attacker may leverage these, we can see the extent to which an attacker can `hide' in noise. Any other interference in the quantum transmissions will raise the noise floor sufficiently that it goes above some calibrated value for $\epsilon$, which would invalidate the proof for $V$.

The argument presented here is congruous with how QKD protocols add security using quantum states. The quantum channel, as with other quantum communications protocols \cite{Cacciapuoti2020}, offers some significant added protection along with the classical security. 

Note, because the security relies on the statistical likelihood of zero measurements, the protocol is fail safe for sufficiently high values of $\epsilon$ above a predetermined noise value from the communication channel.

\subsection{Soundness and ZK}

There are two conditions that ZKPs must aspire to:
\begin{itemize}
\item soundness - that $P$ can only convince $V$ if they really do know a given $x$ and behave honestly, except for some small probability.
\item zero-knowledge - that an neither $V$ nor an eavesdropper $E$ can learn anything about the secret $x$.
\end{itemize}
Both of these follow naturally from the details in section \ref{sec:security}.

Soundness follows directly from the limits given in theorems \ref{thm:classical} and \ref{thm:quantum}, specifically that the only reliable way to attain the correct measurements within error tolerances is for $P$ to provide an honest proof.

Similarly, owing to the structure of $s$ in relation to $p$ and the minimal number of quantum communications from which any value of $x$ could be estimated, the zero-knowledge condition is satisfied.

\subsection{Remarks}

\subsubsection{Mutual ZKP}

Future developments may involve developing the protocol and extending it slightly such that both parties can verify each other - take the challenge committed to in $n$ by $V$. With the addition of another $c_2$ term from $P$, $P$ could also validate $V$ concurrently for the potential of mutual authentication.

\subsubsection{Hardware}

There are several constraints on current hardware that would preclude this from being immediately practical. Namely, the need for a very high precision on the qubit in use, and a likewise minimal amount of noise required to not skew the results. 

Error corrected qubits and quantum communication channels are required to deal with the second part of these issues \cite{Pirandola2015}. The resolution of the qubits and their longevity is taken into account by some benchmarks, such as `Quantum Volume' \cite{Cross2019}. Therefore, as quantum computers grow in reliability and complexity, and quantum networks begin to be tested and deployed and improve, we might consider such high enough resolutions, error correction, and reliability to one day be attainable. 

\section{Conclusion}

This paper hopes to have shown that there is another possibility for performing zero-knowledge proofs using quantum algorithms over quantum communications networks. The protocol in this paper has shown a method to swap out the use of a generator $g$ in Schnorr's scheme for a qubit rotation, and the extra steps required to make a zero-knowledge proof work with currently available algorithms. This system has been shown to have some additional benefits over purely classical approaches, despite its classical origins. 

This work thereby adds to the collection of proposals for QIA and quantum zero-knowledge proofs that might help shape future quantum communications.

\section{Acknowledgements}

The author is thankful to the indulgence of discussion, expertise, and time from Dr. Joseph Wilson and Prof. Ben Varcoe, and to Christoph Graebnitz for identifying a major issue that lead to a redesign of the method in section \ref{sec:QZKP}.

\bibliographystyle{abbrv}
\bibliography{bib}

%\pagebreak
%\appendix

%\section{Appendix}\label{sec:openqasm}
%
%\begin{verbatim}
%OPENQASM 2.0;
%include "qelib1.inc";
%
%qreg q[3];
%creg c[3];
%
%\end{verbatim}

\end{document}